\documentclass[aps,prl,groupedaddress,floats,reprint]{revtex4-1}

\usepackage[english]{babel}

\usepackage{amsfonts,graphicx,soul,natbib,mathrsfs}

\usepackage[colorlinks=true]{hyperref}

\usepackage{bm}
\usepackage{epstopdf}
\usepackage{textcomp}
\usepackage{xcolor}
\usepackage{soul}
\usepackage{physics}
\usepackage{graphics}
\usepackage{comment}
\usepackage{amssymb}

\usepackage{dcolumn}
\usepackage{bm}

\begin{document}

\title{Conversion of twistedness from light to atoms}

\author{S. S. Baturin}%
\affiliation{School of Physics and Engineering,
ITMO University, 197101 St. Petersburg, Russia}%

\author{A. V. Volotka}%
\affiliation{School of Physics and Engineering,
ITMO University, 197101 St. Petersburg, Russia}%

\date{\today}

%=====================================================================
\begin{abstract}
We develop a simple model and propose a scheme that allows the production of twisted atoms in free space using the
absorption of twisted photons by a bound electron. We show that in the inelastic collision of a
photon and an atom, the twisted state of the photon is transferred to the center-of-mass state, so that the projection of the orbital momentum of the atom becomes $m_\gamma-\Delta m_e$. We also show that, depending on the experimental conditions, the twistedness of the photon is either transferred to the atomic center-of-mass quantum state or modifies
the selection rule for the bound electron transition. Proposed scheme is general and enables complex shaping of the atomic wavefront.
\end{abstract}

%=====================================================================

\maketitle

%=====================================================================
{\it Introduction.--}
%=====================================================================
Structured waves with a nonzero topological charge (TC) are actively studied 
in many modern fields of physics. Ranging from acoustics \cite{acc0,acc1,acc2}, elastic waves \cite{elastic1,elastic2}, and hydrodynamics \cite{hydro} to quantum optics \cite{Allen1999,FrankeArnold2008}, high-energy physics \cite{Ivanov2011,IvanovRev}, and atomic physics \cite{Hayrapetyan_2013,Matula_2013,surzhykov2015,afanasev2018,ct6}, this topic attracts more and more researchers and is proven to be very fruitful.

The concept of a wave carrying a TC or possessing a dislocation on the wavefront was introduced and studied by Nye and Berry back in 1974 \cite{Berry1974} and gave rise to a wide modern field of singular optics \cite{Allen1,Allen1999,FrankeArnold2008,Mono}. The next breakthrough in this field is connected to the simple but nontrivial discovery that elementary particles can also carry a TC \cite{Bliokh2007,Lloyd2017,BLIOKH20171} (due to the principle of wave-particle duality). Such particles are usually called \textit{twisted particles}, since the de Broglie wavefront has a helical twist. The TC of twisted particles can be observed as an increase in the projection of the total angular momentum (TAM) on the propagation axis due to the nonzero value of the \textit{intrinsic} orbital angular momentum (OAM) even in free space. For example, in the case of free twisted electrons, the TAM projection can exceed the commonly accepted value of $1/2$ by many times \cite{BLIOKH20171,Lloyd2017}, which is experimentally confirmed \cite{elvort1,elvort2,elvort3}. More details on recent developments can be found in the following reviews \cite{BLIOKH20171,UFN,IvanovRev} and roadmaps \cite{rmap,rmap2}.

The twisted wave concept extends even further and applies to more complex quantum systems such as atoms and molecules \cite{Helseth2004}. For example, coherent atomic beams can experience quantum interference in a computer-generated hologram \cite{CGH,Lembessis2014} and acquire a TC. This idea \cite{Lembessis2014} has been proved in experiment just recently.
\cite{VortAt2021}.

Once the twisted state framework was extended to complex (composite) quantum systems, several fundamental questions immediately appeared. In what sense is a complex system twisted? Is this twist redistributed among the internal degrees of freedom in the process that induces such a state?  A straightforward answer to the first question is to attribute the twist to the center of mass of the system, which seems natural from the perspective of the density matrix formalism. The second question is less obvious, and the answer strongly depends on the experimental conditions. If one assumes that the center of mass of the system cannot move and cannot acquire any motion properties such as momentum and angular momentum, as well as cannot change its energy, then the twistedness can only be accepted by the part that interacts with the twisted species. 
For example, if a twisted photon is absorbed by a bound electron in a trapped atom, the selection rules are modified \cite{Hayrapetyan_2013,Har2,Matula_2013,Davis_2013,scholz-marggraf:2014:013425,surzhykov2015,afanasev2018,afanasev2020,schulz2020}. This striking discovery was experimentally confirmed \cite{schmiegelow2016,Solyanik-Gorgone}. The latest investigations \cite{stopp2022,peshkov2023} show that the modification of the selection rule is not always the case, and once the restriction on the motion of the center of mass is removed, the TC that modifies the TAM of the incoming photon can be shared between the center of mass and the excited electron.  

In this letter, extending this observation, we develop a simple model and propose a scheme that allows the production of twisted atoms and atoms with even more complex structure of the wave function in free space by means of absorption of structured photons by a bound electron. 
%We show that in the inelastic collision of a photon and an atom, the twisted state of the photon is transferred to the center-of-mass state of the atom so that the projection of the orbital momentum of the atomic center of mass becomes $m_\gamma-\Delta m$. Here, $m_\gamma$ is the photon projection of the TAM and $\Delta m$ is the change in the TAM projection of the electron due to its
%transition between bound states. 

Similar studies have been performed for twisted light in the dipole and quadruple approximation in references.\cite{Enk1,Enk2,Jau,Afon,Mond,LightandAtom}. In contrast to the previous results, we explicitly show that it is not only the TAM of the photon that is transferred to the center of mass, but a quantum state, i.e. the transverse structure of the photon is imprinted on the center of mass, allowing complex shaping of the atomic wavefront.

%=====================================================================
{\it Theoretical formulation. --}
%=====================================================================
%
To properly describe a twisted atom, we begin with the Schr\"odinger equation in an inertial coordinate system. The nonrelativistic Hamiltonian of a nucleus with charge number $Z$ and nuclear mass $M$ located at ${\bf r}_n$ and an electron of mass $m$ located at ${\bf r}_e$ interacting with quantized electromagnetic potential takes the following form:
\begin{eqnarray}
\label{eq:mainH}
\hat{\mathcal{H}} &=& \frac{\left[\hat{\mathbf{p}}_e - e\hat{\mathbf{A}}(t,{\bf r}_e)\right]^2}{2m} + V(|\mathbf{r}_n-\mathbf{r}_e|)\nonumber\\
&+& \frac{\left[\hat{\mathbf{p}}_n + eZ\hat{\mathbf{A}}(t,{\bf r}_n)\right]^2}{2M},
\end{eqnarray}
where $V$ is the electron-nucleus interaction potential and $\hat{\mathbf{A}}(t, {\bf r})$ is the (transverse) vector potential of the quantized electromagnetic fields. To simplify the solution, we employ the well-known transformation to the center-of-mass (${\bf R}$) and relative (${\bf r}$) coordinates \cite{bethe}:
\begin{equation}
\label{eq:coord}
\mathbf{R} = \frac{\mathbf{r}_em+\mathbf{r}_nM}{m+M}
\hspace{0.5cm}\text{and}\hspace{0.5cm}
\mathbf{r} = \mathbf{r}_e - \mathbf{r}_n.
\end{equation}
The momenta transform as follows 
\begin{equation}
\label{eq:mom}
\hat{\mathbf{P}} = \hat{\mathbf{p}}_e + \hat{\mathbf{p}}_n
\hspace{0.5cm}\text{and}\hspace{0.5cm}
\hat{\mathbf{p}} = \hat{\mathbf{p}}_e - \frac{m}{m+M} \left(\hat{\mathbf{p}}_e + \hat{\mathbf{p}}_n \right).
\end{equation}
Substituting the above equations into Eq.~\eqref{eq:mainH} and keeping only
the lowest order in $m/M$, we get 
\begin{align}
\hat{\mathcal{H}} = \hat{\mathcal{H}}_0 + \hat{\mathcal{H}}_i + \mathcal{O}\left[\frac{m}{M} \right]
\end{align}
with the unperturbed Hamiltonian $\hat{\mathcal{H}}_0$,
\begin{align}
\label{eq:H0}
    \hat{\mathcal{H}}_0 = \frac{\hat{\mathbf{P}}^2}{2M} + \frac{\hat{\mathbf{p}}^2}{2m} + V(r),
\end{align}
and the interaction part $\hat{\mathcal{H}}_i$,
\begin{equation}
\label{eq:Hint}
\hat{\mathcal{H}}_i = \frac{eZ}{M}\hat{\mathbf{P}}\hat{\mathbf{A}}(t,\mathbf{R}) - \frac{e}{M}\hat{\mathbf{P}}\hat{\mathbf{A}}(t,{\bf R+r})
- \frac{e}{m}\hat{\mathbf{p}}\hat{\mathbf{A}}(t,{\bf R+r}).   
\end{equation}
Here, we keep only the terms linear in $\hat{\mathbf{A}}$, since 
we consider one-photon process of atomic photoionization. Due to the separation of variables, the eigenfunctions of the zero-order Hamiltonian are now easy to find:  
\begin{equation}
\label{eq:wavefactor}
    \hat{\mathcal{H}}_0 \Phi(t,{\bf R}) \phi(t,{\bf r}) = (E+\varepsilon)\Phi(t,{\bf R}) \phi(t,{\bf r}),
\end{equation}
where $\Phi$ and $\phi$ are the wavefunctions of the freely propagating center of mass and the bound electron, respectively, with the corresponding eigenvalues $E$ and $\varepsilon$.

%=====================================================================
{\it Absorption of the photon. -- }
%=====================================================================
%
We now consider the process of photo-absorption by an atom:
\begin{equation}
    A({\bf P}_a, \varepsilon_a) + f \rightarrow B({\bf P}_b, \varepsilon_b),
\end{equation}
where the initial/final state of the
atom $A$/$B$ is described by the momentum of the center of mass ${\bf P}_a$/${\bf P}_b$ and by the energy of the bound electron $\varepsilon_a$/$\varepsilon_b$, and $f$ indicates the 
quantum numbers of the photon. The corresponding $S$-matrix element reads \cite{BLP}
\begin{eqnarray}
\label{eq:S0}
S_{ba} &\equiv& \langle {\bf P}_b, \varepsilon_b | S | {\bf P}_a, \varepsilon_a, f \rangle = -i \int_{-\infty}^\infty dt \int d^3 R\, d^3 r\nonumber\\
&\times&\Phi^*_{{\bf P}_b}(t,{\bf R})\phi^*_{b}(t,{\bf r}) \langle 0 | \hat{\mathcal{H}}_i | f \rangle \Phi_{{\bf P}_a}(t,{\bf R})\phi_{a}(t,{\bf r}),
\end{eqnarray}
where $\Phi_{{\bf P}_a}$ and $\Phi_{{\bf P}_b}$ are the plane-wave solutions of the equation of motion of the center of mass 
\begin{equation}
\Phi_{{\bf P}_a}(t,{\bf R}) = \exp[-i(E_a t - {\bf P}_a {\bf R})]
\end{equation}
with $E_a = P_a^2/(2M)$. Expanding the quantized electromagnetic potential in terms of creation and annihilation operators, we obtain
\begin{eqnarray}
\label{eq:Hint1}
\langle 0 | \hat{\mathcal{H}}_i | f \rangle &=& \frac{eZ}{M}\hat{\mathbf{P}}\mathbf{A}_f(t,\mathbf{R}) - \frac{e}{M}\hat{\mathbf{P}}\mathbf{A}_f(t,{\bf R+r})\nonumber\\
&-&\frac{e}{m}\hat{\mathbf{p}}\mathbf{A}_f(t,{\bf R+r}),
\end{eqnarray}
where ${\bf A}_f$ is the photon wave function, ${\bf A}_f(t,{\bf r}) = \exp(-i\omega t) {\bf A}_f({\bf r})$, and $\omega$ is the photon energy. Here, one can neglect the contributions of the first and the second terms in Eq.~\eqref{eq:Hint1}, since in the long-wavelength approximation, they correspond to the photon absorption by the center of mass $[e(Z-1)/M] \hat{\mathbf{P}}\mathbf{A}_f({\bf R})$ without the change in 
the electron configuration. Further,
higher-multipole contributions of the second term should be considered together with the relativistic correction \cite{au:1997:162, barton:2005:521}. Thus, we omit the 
terms with ${\bf P}$, and the $S$-matrix takes the
form
\begin{eqnarray}
\label{eq:S1}
S_{ba}&=& 2\pi i \delta(E_a+\varepsilon_a+\omega-E_b-\varepsilon_b) \int d^3 R\, d^3 r\nonumber\\
&\times&\Phi^*_{{\bf P}_b}({\bf R}) \phi^*_{b}({\bf r}) \frac{e}{m} \hat{{\bf p}} {\bf A}_f({\bf R+r})\Phi_{{\bf P}_a}({\bf R})\phi_{a}({\bf r}).
\end{eqnarray}

\begin{figure}[t]
    \centering
     \includegraphics[width=0.49\textwidth]{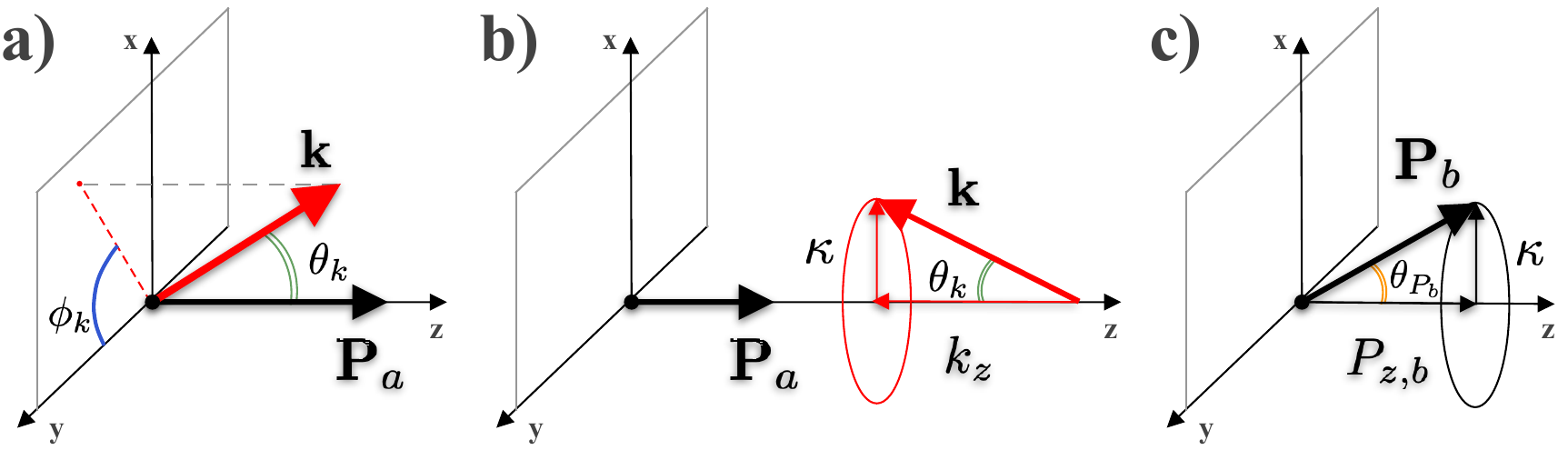}
    \caption{Schematics of the momentum vector for different cases. %Panel
    a) %-
    general plane wave case; %panel
    b) %-
    a twisted photon and a plane wave center of mass; %panel
    c) %-
    twisted final center of mass. 
    %{\color{red} $\theta_k -> \theta_{P_{\perp,b}}$}
    }
    \label{Fig:0}
\end{figure}
%=====================================================================
{\it Final state of the center of mass. --}
%=====================================================================
We approach the main 
question of
the paper: 
which state does the center of mass evolve to after the photo-absorption? According to the formalism of the $S$, the final state can be found as \cite{BLP}:
\begin{align}
|\rm{out} \rangle = \sum\limits_{\alpha}|\alpha\rangle \langle \alpha| \hat{S}|\rm{in} \rangle,
\end{align}
where the summation is performed
over the entire Hilbert space of final states $\alpha$. For
convenience purposes, we use the plane wave basis set.  
Thus, the final state of the center of mass $\Phi_b(t,{\bf R})$ is:
\begin{align}
\label{eq:wf}
\Phi_b(t,{\bf R}) = \int \Phi_{{\bf P}_b}(t,{\bf R}) S_{ba} \frac{d^3 P_b}{(2\pi)^3}.
\end{align}
To proceed further, one has to specify the state of the absorbed photon.

%=====================================================================
{\it Plane-wave photon. --}
%=====================================================================
%
When the absorbed photon is described by a
plane wave with the wave-vector ${\bf k}$ and polarization vector $\boldsymbol{\epsilon}_{{\bf k}\Lambda}$, the coordinate part of the photon wavefunction reads
\begin{align}
\label{eq:APW}
\mathbf{A}_{f}(\mathbf{R}+\mathbf{r}) \equiv \mathbf{A}_{{\bf k}\Lambda}(\mathbf{R}+\mathbf{r}) = \frac{\boldsymbol{\epsilon}_{{\bf k}\Lambda}}{\sqrt{2\omega}}\exp[i\mathbf{k}(\mathbf{R}+\mathbf{r})].
\end{align}
Substituting \eqref{eq:APW} into Eq.~\eqref{eq:S1}, we arrive at
\begin{eqnarray}
S_{ba}^{\rm PW} &=& (2\pi)^3 \delta(E_a + \varepsilon_a + \omega - E_b - \varepsilon_b) \nonumber\\
&\times&\delta({\bf P}_a + {\bf k} - {\bf P}_b) {\cal M}_{m_b m_a}(\theta_k,\varphi_k),
\end{eqnarray}
with the electronic transition matrix element ${\cal M}_{m_b m_a}(\theta_k,\varphi_k)$ defined as
\begin{eqnarray}
\label{eq:matelem1}
{\cal M}_{m_b m_a}(\theta_k,\varphi_k) &=& \frac{2\pi i}{\sqrt{2\omega}} \frac{e}{m}\int d^3r\,\phi^*_{n_b l_b m_b}({\bf r}) \nonumber\\
&\times&\exp(i\mathbf{k}\mathbf{r}) \boldsymbol{\epsilon}_{{\bf k}\Lambda} \hat{{\bf p}}\, \phi_{n_a l_a m_a}({\bf r}),   
\end{eqnarray}
where $n_a$, $l_a$, $m_a$ ($n_b$, $l_b$, $m_b$) are the quantum numbers of the initial (final) state of the bound electron and $(\theta_k, \varphi_k)$ are the polar and azimuthal components of the wave vector ${\bf k}$, shown 
in Fig. 1a. Using Eq.~\eqref{eq:wf}, we obtain
\begin{eqnarray}
\label{eq:WFPW}
    &\Phi_b^{\rm PW}&(t,{\bf R}) = \exp[-iE_bt + i({\bf P}_a + {\bf k}){\bf R}]\nonumber\\
    &\times&\delta(E_a + \varepsilon_a + \omega - E_b - \varepsilon_b){\cal M}_{m_b m_a}(\theta_k,\varphi_k).
\end{eqnarray}
Thus, the final state of the center of mass after the absorption of the plane-wave photon is a
{\it plane wave} with a well-defined energy $E_b = ({\bf P}_a + {\bf k})^2/(2M)$ and momentum ${\bf P}_b ={ \bf P}_a + {\bf k}$. The $\delta$-function fixes the photon energy by the resonance condition $E_a + \varepsilon_a + \omega - E_b - \varepsilon_b = 0$, and ${\cal M}_{m_b m_a}(\theta_k,\varphi_k)$ characterizes the amplitude of the process.

%=====================================================================
{\it Twisted-wave photon. --}
%=====================================================================
%
Next, we consider the interaction of an atom with twisted radiation. We propose a collinear scenario: the $z$- axis (quantization 
axis) is chosen along the initial center-of-mass momentum ${\bf P}_a = (0,0,P_{z,a})$, and the twisted photon wave packet propagates along the $z$-axis towards the atom, see Fig. \ref{Fig:1}b. We note that non-zero impact parameters and non-collinear scenarios do not, in principle, affect the concept. We provide this analysis in the supplementary materials, but retain a simplified model.

In the collinear case, the wave function of the twisted photon ${\bf A}_{k_z \kappa m_\gamma \Lambda}$ can be written as a superposition of the plane waves in the following
form \cite{jentschura:2011:013001, Jentschura:2011aa}:
\begin{equation}
\label{eq:atw}
{\bf A}_{k_z \kappa m_\gamma \Lambda}({\bf R+r}) = \int \boldsymbol{\epsilon}_{{\bf k}\Lambda} e^{i{\bf k}({\bf R+ r})} a_{\kappa, m_\gamma}({\bf k}_\perp) \frac{d^2 k_\perp}{(2\pi)^2}
\end{equation}
with the amplitude
\begin{equation}
a_{\kappa, m_\gamma}({\bf k}_\perp) = (-i)^{m_\gamma} e^{im_\gamma\varphi_k} \sqrt{\frac{2\pi}{k_\perp}} \delta(k_\perp-\kappa).
\end{equation}
The following quantum numbers are well-defined for the twisted photon: $k_z$ is the longitudinal momentum, $\kappa$ is the absolute value of the transverse momentum, $m_\gamma$ is the projection of the total angular momentum (TAM) on the propagation axis, and $\Lambda$ is the photon helicity. One can also define the opening angle $\theta_k$ as $\theta_k = \kappa/k_z$. By substituting Eq.~\eqref{eq:atw} into Eq.~\eqref{eq:S1} and integrating plane waves, we arrive at
\begin{align}
\label{eq:STW}
&S_{ba}^{\rm TW} = (2\pi)^3 \delta(E_a + \varepsilon_a + \omega - E_b - \varepsilon_b) \delta(P_{z,a} - k_z - P_{z,b})\nonumber\\
&\times\int \delta({\bf k}_\perp - {\bf P}_{\perp,b}) a_{\kappa, m_\gamma}({\bf k}_\perp) {\cal M}_{m_b m_a}(\theta_k,\phi_k) \frac{d^2 k_\perp}{(2\pi)^2}.
\end{align}
We note the transverse $\delta$-function and immediately conclude that the center of mass acquires the perpendicular momentum of the photon. To investigate the phase, we have to identify the dependence of the the electronic transition matrix element ${\cal M}_{m_b m_a}(\theta_k,\varphi_k)$ on the azimuthal angle $\varphi_k$. This can be done with the help of Wigner $D$-functions, as was demonstrated in Ref.~\cite{scholz-marggraf:2014:013425}. The electron wavefunction can be rotated to the
quantization axis along ${\bf k}$, by two Euler angles $\theta_k$ around the $y$ axis and an angle $\phi_k$ around the $z$ axis, as follows \cite{varshalovich}
\begin{align}
\phi_{nlm}(\mathbf{r}) = \sum_{m'}D^{l*}_{mm'}(\varphi_k,\theta_k,0)\phi_{nlm'}(\mathbf{r}')
\end{align}
where $D^{l}_{mm'}(\varphi_k,\theta_k,0) = e^{-im\varphi_k} d^{l}_{mm'}(\theta_k)$ is the Wigner $D$ function. The electronic transition matrix element can be then written as
\begin{eqnarray}
\label{eq:M}
{\cal M}_{m_b m_a}(\theta_k,\varphi_k) &=& \sum_{m_a'm_b'} e^{i(m_a-m_b)\varphi_k} d^{l_b}_{m_b m_b'}(\theta_k) d^{l_a}_{m_a m_a'}(\theta_k)\nonumber\\
&\times&{\cal M}_{m_b' m_a'}(0,0),
\end{eqnarray}
with the collinear
matrix element 
\begin{eqnarray}
{\cal M}_{m_b' m_a'}(0,0) &=& \frac{2\pi i}{\sqrt{2\omega}} \frac{e}{m}\int d^3r'\,\phi^*_{n_b l_b m_b'}({\bf r'}) \nonumber\\
&\times&\exp(ikz') \hat{p}_\Lambda\, \phi_{n_a l_a m_a'}({\bf r'}),
\end{eqnarray}
where $p_\Lambda = -i\nabla_\Lambda$ is the cyclic ($\Lambda = \pm 1$) projection of the momentum operator. Substituting Eq.~\eqref{eq:M} into Eq.~\eqref{eq:STW} and then into Eq.~\eqref{eq:wf}, we get
%==========================================================
\begin{figure}[t]
    \centering
     \includegraphics[width=0.49\textwidth]{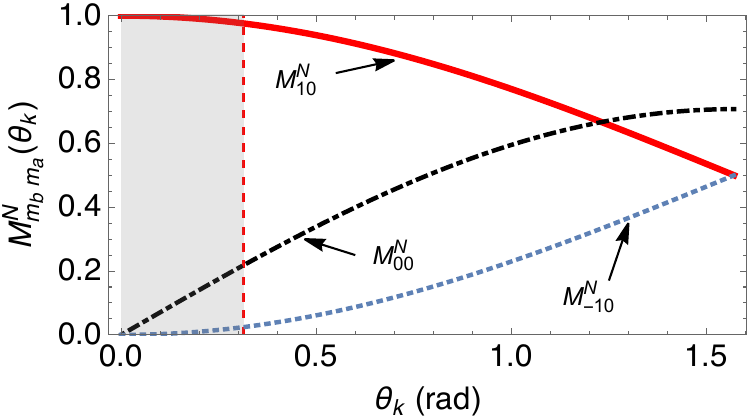}
    \caption{Normalized amplitude of the process $M^N_{m_b m_a}=\widetilde{{\cal M}}_{m_b m_a}(\theta_k)/\widetilde{{\cal M}}_{1 0}(0)$ as a function of the photon opening angle $\theta_k$: for the case of $\Lambda=1$, $m_a=0$ and $m_b=1$, shown with
    solid red line; for $m_b=0$,
    dot-dashed black line; and for $m_b=-1$, dashed blue line. Grey area highlights the range of $\theta_k\in [0,\pi/10]$, where the transition with $m_b=1$ is dominant. 
    }
    \label{Fig:3}
\end{figure}
%==================================================================
\begin{eqnarray}
\label{eq:WFTW}
&\Phi_b^{\rm TW}&(t,{\bf R}) = \exp[-iE_bt+i(P_{z,a}-k_z)z]\nonumber\\
&\times&\int e^{i{\bf P}_{\perp,b}\mathbf{R}} a_{\kappa, m_a+m_\gamma-m_b}({\bf P}_{\perp,b}) \frac{d^2P_{\perp,b}}{(2\pi)^2} \nonumber\\
&\times&\delta(E_a+\varepsilon_a+\omega-E_b-\varepsilon_b) \widetilde{\cal M}_{m_b m_a}(\theta_k)
\end{eqnarray}
with $E_b = [(P_{z,a}-k_z)^2 + \kappa^2] / (2M)$,
\begin{eqnarray}
a_{\kappa, m_a+m_\gamma-m_b}({\bf P}_{\perp,b}) &=& (-i)^{m_a+m_\gamma-m_b} e^{i(m_a+m_\gamma-m_b)\varphi_{P_{\perp,b}}}\nonumber\\
&\times&\sqrt{\frac{2\pi}{P_{\perp,b}}} \delta(P_{\perp,b}-\kappa)
\end{eqnarray}
and
\begin{eqnarray}
\widetilde{{\cal M}}_{m_b m_a}(\theta_k) &=& i^{m_a-m_b}\sum_{m_a'm_b'}d^{l_b}_{m_b m_b'}(\theta_k) d^{l_a}_{m_a m_a'}(\theta_k)\nonumber\\
&\times&{\cal M}_{m_b' m_a'}(0,0).
\end{eqnarray}
Eq.~\eqref{eq:WFTW} represents the main result of the paper. The first and second lines characterize the properties of the final state of the center of mass,  
the $\delta$-function ensures the resonance condition, and $\widetilde{\cal M}_{m_b m_a}(\theta_k)$ quantifies the amplitude of the process. It is easy to see that the wave function $\Phi_b^{\rm TW}$ is the eigenstate of the squared momentum operator
\begin{equation}
\hat{\bf P}^2 \Phi_b^{\rm TW}(t,{\bf R}) = [(P_{a,z}-k_z)^2 + \kappa^2] \Phi_b^{\rm TW}(t,{\bf R}) 
\end{equation}
and the longitudinal momentum operator
\begin{equation}
\hat{P}_z \Phi_b^{\rm TW}(t,{\bf R}) = (P_{a,z}-k_z) \Phi_b^{\rm TW}(t,{\bf R}).
\end{equation}
The transverse integral in Eq.~\eqref{eq:WFTW} can be evaluated as \cite{Matula_2013}
\begin{align}
\int e^{i{\bf P}_{\perp,b}\mathbf{R}}& a_{\kappa, m_a+m_\gamma-m_b}({\bf P}_{\perp,b}) \frac{d^2P_{\perp,b}}{(2\pi)^2}\\ \nonumber
&= \sqrt{\frac{\kappa}{2\pi}} J_{m_a+m_\gamma-m_b}(\kappa R_\perp) e^{i(m_a+m_\gamma-m_b)\phi_R}.
\end{align}
From the expression above, it becomes clear that $\Phi_b^{\rm TW}$ is also an eigenfunction of the operator $\hat{L}_z$ 
%of the $z$-projection of the} OAM % OAM $z$-projection operator $\hat{L}_z$
%
\begin{equation}
\hat{L}_z \Phi_b^{\rm TW}(t,{\bf R}) = (m_a+m_\gamma-m_b) \Phi_b^{\rm TW}(t,{\bf R}),
\end{equation}
and, consequently, has a defined OAM projection on the $z$-axis.

Thus, we explicitly demonstrate that the absorption of 
twisted light converts an atom to a {\it twisted state}, see Fig. 1c. Its opening angle $\theta_{P_b}$ is equal to the ratio of the transversal momentum to the longitudinal momentum, $\theta_{P_b} = \arctan[\kappa/(P_{a,z}-k_z)]$. At the same time, the photon's TAM projection is shared by the electron and the center of mass, i.e., $(m_b-m_a)$ goes to the bound electron, while the rest, $m_\gamma - m_b + m_a$, goes to the center of mass.

The amplitude of the process $\widetilde{\cal M}_{m_b m_a}(\theta_k)$ does not depend on the photon's TAM projection and, therefore, the twistedness of the photon does not change the atomic selection rules. Similar to the ordinary plane wave case, the selection rules are determined by the initial and final electronic states, which define the allowed multipoles in the photon wave function expansion. To illustrate this observation, we numerically evaluate the amplitude $\widetilde{\cal M}_{m_b m_a}(\theta_k)$ for a hydrogen-like atom excited from its ground state $1s$ ($m_a = 0$) to the $2p$ ($m_b = 0, \pm 1$) state as a function of angle $\theta_k$. The results are presented in Fig.~\ref{Fig:3}.

In Fig.~\ref{Fig:3}, we show that for $\theta_k \rightarrow 0$ and for $\Lambda = 1$, the only non-vanishing amplitude is 
$\widetilde{\cal M}_{1 0}(\theta_k)$, as expected from the common selection rule. When the photon opening angle increases, the amplitudes with $m_b = 0$ and $m_b = -1$ start to contribute as well. This means that the final state of center of mass becomes a mixture of twisted states with TAM projections $m_\gamma - 1$, $m_\gamma$, and $m_\gamma + 1$. However, for large $m_\gamma$, we can neglect the unity and note that the center of mass is a twisted state with an almost defined TAM projection (with a small dispersion of TAM projection). Moreover, as one can see from  Fig.~\ref{Fig:3}, for relatively small photon opening angles, i.e. $\theta_k \le \pi/10$, the dominant channel is $m_b = 1$. The latter  ensures that the center-of-mass TAM projection is equal to 
$m_\gamma - 1$ for the considered case.

Another idea that
can be implemented experimentally is to place
an atom in a constant magnetic field. In this case, the Zeeman sublevels will be energetically resolved. Then, one can choose the photon energy equal to the transition between the ground electron state and the magnetic sublevel $m_b$. In this scenario, the final state of the center of mass will be a pure twisted state with a TAM projection of $m_\gamma + m_a - m_b$, as other excitation channels are forbidden. 
The possible experimental scheme is presented in Fig.~\ref{Fig:1}: an atomic (ion) beam passes through a solenoid where it intersects with the twisted laser beam.
\begin{figure}[t]
    \centering
     \includegraphics[width=0.5\textwidth]{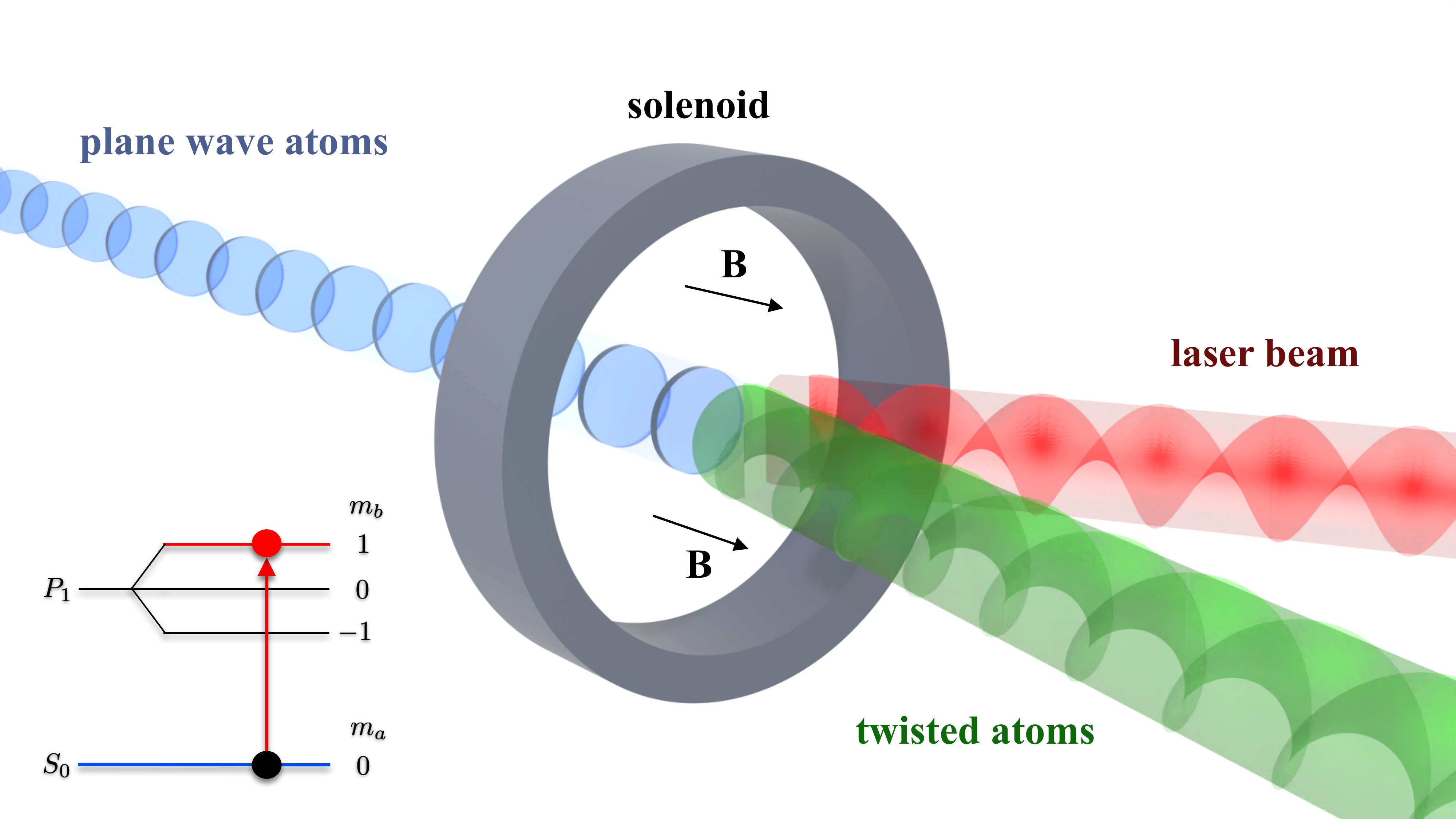}
    \caption{Experimental scheme for conversion of TAM from photon to atom in a 
    magnetic field of a solenoid. 
    Due to the Zeeman splitting, the ground state ($S_0$ state) of an atom can be selectively excited into $P_1$ state with projection +1. At the output, the atomic beam (center-of-mass state) gains TAM projection $m_\gamma - 1$ and evolves to a     twisted state.
    %{\color{red} $1s$ to $S_0$ and $2p$ to $P_1$; $S$ should not be splitted; for p-state indicate the projection from the right side}
    }
    \label{Fig:1}
\end{figure}
%

%=====================================================================
{\it Limit of infinite %ly heavy center of 
mass. --}
%=====================================================================
%
It is instructive to consider the limit of $M \rightarrow \infty$. In this case, we assume that both ${\bf P}_a$ and ${\bf P}_b$ are much larger than ${\bf k}$, and the center of mass
does not change its momentum ${\bf P}_b = {\bf P}_b$. In turn, it means that ${\bf P}_{\perp,b} = 0$, and one can easily perform the integration over $\varphi_{P_{\perp,b}}$ in 
Eq.~\eqref{eq:WFTW}. The final state of the center of mass in the limit of infinite mass reads
\begin{eqnarray}
\label{eq:limit}
 \Phi^{\rm TW}_{b}(t,{\bf R})\Big|_{M \rightarrow \infty} &\propto& \exp[-i E_a t + i P_{z,a}z] \widetilde{\cal M}_{m_b m_a}(\theta_k)\nonumber\\
 &\times& \delta_{m_\gamma, m_b-m_a} \delta(\varepsilon_a + \omega - \varepsilon_b).
\end{eqnarray}
Evidently, the Kronecker $\delta$-symbol ensures that $m_\gamma = m_b - m_a$, 
meaning that the twisted light with TAM projection $m_\gamma$ can excite only the transitions with $\Delta m = m_\gamma$. This {\it modified selection rule} was earlier obtained in various theoretical works \cite{Davis_2013, scholz-marggraf:2014:013425, surzhykov2015, Solyanik-Gorgone, afanasev2020} and confirmed in experiments with photo-absorption of twisted light by an ion in the Paul trap \cite{schmiegelow2016, afanasev2018}. In measurement scheme, the condition $M \rightarrow \infty$ was realized by the trap potential, where the motion of the center of mass is quantized to vibration levels. 
Once the ion is cooled to the lowest vibration level, only resonant excitation of the electronic degree of freedom is
possible, as the transfer of the photon's energy and TAM to the center of mass is energetically forbidden. In a  followup 
experiment \cite{stopp2022}, however, Stopp {\it et al.} demonstrated that when excitation of the sideband structure of the ion motion is allowed, the photon's TAM can be shared between the bound electron and the center of mass, which has also been
described theoretically recently \cite{peshkov2023}.

%=====================================================================
{\it Conclusion. --}
%=====================================================================
%
We have considered the interaction of twisted light with 
a composite system within the one-electron atom (ion) model. 
The presented formulation can be easily extended to a
many-electron system for investigation of 
more complex interactions. 
We have shown how the total angular momentum (TAM) of a
photon can be redistributed between the center-of-mass and the bound-electron degrees of freedom in full agreement with the previous studies \cite{Enk1,Enk2,Jau,Afon,Mond,LightandAtom}.
 %The previously found modified selection rules are no longer applicable in case when the photon's TAM is transferred to the center-of-mass motion, and regular selection rules are recovered. 
 We have revealed that for a collinear scenario,
the structure of the photon 's wave function as well as TAM is completely transferred to the atom, and the atomic wave function becomes twisted. 
Since the cross-section of such a resonance process is rather large, it can be used to produce twisted atomic (ionic) beams. In addition, this technique can be used to create structured atomic beams of various shapes, such as Hermitt-Gauss, Airy, and others.
Such shaping of the atomic wavefront can be very useful in the scattering process with twisted atoms, as it could help to extract new types of information the same way it as with the help of twisted neutrons \cite{afanasev:2019:051601}. 
Moreover, collisions of ions in a vortex state are expected to exhibit different properties compared to usual ones, e.g. 
nuclear reactions could be altered.
%
%=====================================================================
\begin{acknowledgments}

%The work is funded by ....

%Lidiya Pogorelskaya english

%Cover letter

%Russian Science Foundation and St. Peterburg Science Foundation, project № 22-22-20062, https://rscf.ru/project/22-22-20062/. 
%RSF 22-12-00258
%Ministry of Science and Education of the Russian Federation (Project No. 075-15-2021-1349
The authors thank Igor Chestnov, Dmitriy Karlovets, and Ivan Terekhov for useful discussions and suggestions and Lidiya Pogorelskaya for careful reading of the manuscript.
\end{acknowledgments}
%=====================================================================

\begin{widetext}
\section{Photon wave function}
The theoretical foundations of the twisted photon can be found in Ref.\cite{UFN}. The wave function of the twisted photon with a well-defined projection of the orbital angular momentum on the propagation axis can be written as a superposition of Bessel functions:
\begin{align}
\label{eq:phWF}
  {\bf A}_{k_z \kappa m_\gamma \Lambda}({\bf r}) = \sum\limits_{\sigma=0,\pm1} i^{-\sigma} d_{\sigma,\Lambda}(\theta_k)J_{m_\gamma-\sigma} (\kappa r) \exp \left[i (m_\gamma-\sigma) \phi_r \right] \boldsymbol{\chi}_\sigma \exp(i k_z z). 
\end{align}
Above, $d_{\sigma,\Lambda}(\theta_k)$- is the small Wigner function \cite{varshalovich}, $\boldsymbol{\chi}_\sigma$ is the eigenvector of the $z$-projection spin operator $\hat{s}_z$
\begin{align}
    \hat{s}_z \boldsymbol{\chi}_\sigma = \sigma \boldsymbol{\chi}_\sigma; 
\end{align}

with $\sigma=0,\pm1$, and $\Lambda=\pm1$ is the eigenvalue of the helicity operator $\hat{\Lambda}=\hat{\boldsymbol{s}}\boldsymbol{k}/k:$
\begin{align}
    \hat{\Lambda} \boldsymbol{\epsilon}_{{\bf k}\Lambda}=\Lambda \boldsymbol{\epsilon}_{{\bf k}\Lambda}.
\end{align}
Above $\boldsymbol{\epsilon}_{{\bf k}\Lambda}$ is the polarization vector.

Vectors $\boldsymbol{\chi}_\sigma$ have the form:
\begin{align}
\label{eq:chi}
    \bm \chi_{+1}=-\frac{1}{\sqrt{2}}\left(\begin{matrix} 1 \\ i \\ 0 \end{matrix}\right),~~
    \bm \chi_{-1}=\frac{1}{\sqrt{2}}\left(\begin{matrix} 1 \\ -i \\ 0 \end{matrix}\right),~~
    \bm \chi_{0}=\left(\begin{matrix} 0 \\ 0 \\ 1 \end{matrix}\right).
\end{align}

Equation Eq.\eqref{eq:phWF} can be presented in an integral form \cite{UFN,jentschura:2011:013001,Jentschura:2011aa,scholz-marggraf:2014:013425,surzhykov2015,surzhykov:2016:033420} 
\begin{align}
{\bf A}_{k_z \kappa m_\gamma \Lambda}({\bf r}) = \int \boldsymbol{\epsilon}_{{\bf k}\Lambda} e^{i{\bf k}({\bf r})} a_{\kappa, m_\gamma}({\bf k}_\perp) \frac{d^2 k_\perp}{(2\pi)^2}.
\end{align}
Here the amplitude $a_{\kappa, m_\gamma}({\bf k}_\perp)$ is given by
\begin{equation}
a_{\kappa, m_\gamma}({\bf k}_\perp) = (-i)^{m_\gamma} e^{im_\gamma\varphi_k} \sqrt{\frac{2\pi}{k_\perp}} \delta(k_\perp-\kappa).
\end{equation}
The representation above is used in the main article for the calculations of the corresponding $S$-matrix element that describes the interaction with the twisted photon.

To illustrate Eq.\eqref{eq:phWF} we plot the probability density of the photon and the phase distribution in Fig.\ref{Fig:S1}.

\begin{figure*}[t]
    \centering
     \includegraphics[width=0.9\textwidth]{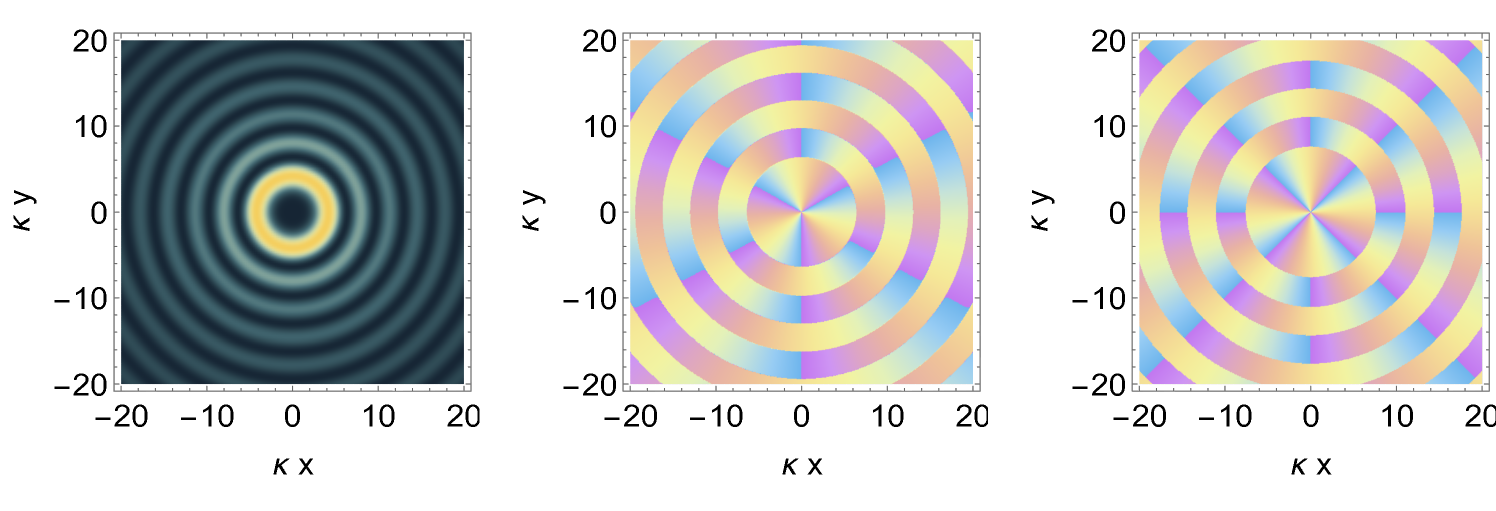}
    \caption{Probability density distribution ${\bf A}^*({\bf r})\cdot {\bf A}({\bf r})$ as a function of the normalized transverse coordinates $\kappa x$ and $\kappa y$ - left panel; phase distribution for the x-component of the photon wave function $\arg[A_x({\bf r})]$ as a function of the normalized transverse coordinates $\kappa x$ and $\kappa y$ middle panel and phase distribution for the z-component of the photon wave function $\arg[A_z({\bf r})]$ as a function of the normalized transverse coordinates $\kappa x$ and $\kappa y$ right panel. The parameter for the plot are $m=4$, $\theta_k=0.2$ and $\Lambda=1$.
    }
    \label{Fig:S1}
\end{figure*}

\section{Nonzero impact parameter}
In the main article, we considered the case where the photon wave packet and the atomic center of mass wave counter-propagate along the same axis. This situation may seem artificial, but it captures the essence of the inelastic scattering mechanism quite well.

Indeed in general the axis of propagation of the photon wave packet can be shifted \cite{scholz-marggraf:2014:013425}. This shift is usually characterized by the vector impact parameter ${\bf b}^{T}=(b_x,b_y,0)$ and can be accounted for as an additional term in the photon wave function as
\begin{align}
\label{eq:MS1}
{\bf A}_{k_z \kappa m_\gamma \Lambda}({\bf R+r-b}) = \int \boldsymbol{\epsilon}_{{\bf k}\Lambda} e^{i{\bf k}({\bf R+r-b})} a_{\kappa, m_\gamma}({\bf k}_\perp) \frac{d^2 k_\perp}{(2\pi)^2}.
\end{align}
It is clear that the substitution 
\begin{align}
    {\bf \Tilde{R}=R-b}
\end{align}
Reduces Eq.\eqref{eq:MS1} to the Eq.(19) of the main text.

Thus, the wave function of the center of mass after the photon absorption has the form
\begin{eqnarray}
\label{eq:WFTW}
&\Phi_b^{\rm TW}&(t,{\bf \tilde{R}}) = \exp[-iE_bt+i(P_{z,a}-k_z)z]
\sqrt{\frac{\kappa}{2\pi}} J_{m_a+m_\gamma-m_b}(\kappa \tilde{R}_\perp) e^{i(m_a+m_\gamma-m_b)\phi_{\tilde{R}_\perp}} \nonumber\\
&\times&\delta(E_a+\varepsilon_a+\omega-E_b-\varepsilon_b) \widetilde{\cal M}_{m_b m_a}(\theta_k)
\end{eqnarray}
It is immediately evident that the final state of the center of mass is twisted with the well-defined projection of the TAM on the $\tilde{z}$-axis, that coincide with the the initial propagation axis of the twisted photon. 

\section{Nonzero crossing angle}

In a general noncollinear case the $S$ matrix element of the twisted photon absorption by the electron in the atom reads
\begin{align}
\label{eq:STWS1}
&S_{ba}^{\rm TW} = (2\pi)^3 \delta(E_a + \varepsilon_a + \omega - E_b - \varepsilon_b) \delta(P_{z,a} - k_z - P_{z,b})\nonumber\\
&\times\int \delta({\bf P}_{\perp,a}+{\bf k}_\perp - {\bf P}_{\perp,b}) a_{\kappa, m_\gamma}({\bf k}_\perp) {\cal M}_{m_b m_a}(\theta_k,\phi_k) \frac{d^2 k_\perp}{(2\pi)^2}.
\end{align}
The expression above differs from the corresponding matrix element in the main article by an additional term ${\bf P}_{\perp,a}$ in the transverse delta function under the integral. 

With the Eq.\eqref{eq:STWS1} expression for the wave function of the center of mass after the photon absorption modifies to
\begin{eqnarray}
\label{eq:WFTWmod}
&\Phi_b^{\rm TW}&(t,{\bf R}) = \exp[-iE_bt+i(P_{z,a}-k_z)z]
\int e^{i{\bf P}_{\perp,b}\mathbf{R}} a_{\kappa, m_a+m_\gamma-m_b}({\bf P}_{\perp,b}-{\bf P}_{\perp,a}) \frac{d^2P_{\perp,b}}{(2\pi)^2} \nonumber\\
&\times&\delta(E_a+\varepsilon_a+\omega-E_b-\varepsilon_b) \widetilde{\cal M}_{m_b m_a}(\theta_k).
\end{eqnarray}
Since the integration is taken over the entire range $P_{x,b}\in [-\infty,\infty]$ and $P_{y,b}\in [-\infty,\infty]$ we may introduce coordinate substitution 
\begin{align}
    \tilde{{\bf P}}_{\perp,b}={\bf P}_{\perp,b}-{\bf P}_{\perp,a}
\end{align}
and get
\begin{eqnarray}
\label{eq:WFTWmod3}
&\Phi_b^{\rm TW}&(t,{\bf R}) = \exp[-iE_bt+i(P_{z,a}-k_z)z]\exp[i {\bf P}_{\perp,a}{\bf R}]
\int e^{i\tilde{{\bf P}}_{\perp,b}\mathbf{R}} a_{\kappa, m_a+m_\gamma-m_b}(\tilde{{\bf P}}_{\perp,b}) \frac{d^2\tilde{P}_{\perp,b}}{(2\pi)^2} \nonumber\\
&\times&\delta(E_a+\varepsilon_a+\omega-E_b-\varepsilon_b) \widetilde{\cal M}_{m_b m_a}(\theta_k).
\end{eqnarray}
We can see that Eq.\eqref{eq:WFTWmod3} is identical to Eq. (25) of the main text, with the exception of an additional phase factor $\exp[i {\bf P}_{\perp,a}{\bf R}]$. This term tilts the phase, which means that the final state is no longer a pure state. However, the mean value of the TAM projection on the propagation axis of the center of mass state is still equal to the $m_a+m_\gamma-m_b$, which means that the strength of the phase singularity or the topological charge of the center of mass quantum state remains the same despite the noncollinearity of the process. A straightforward evaluation of the topological charge \cite{Berry1974,Berry2} yields
\begin{align}
\label{eq:charge}
    m_a+m_\gamma-m_b=\lim\limits_{R_\perp\to \infty}  \frac{1}{2\pi} \int\limits_{0}^{2\pi}d\varphi_R \frac{d \left(\arg \Phi_b^{\rm TW} \right)}{d\varphi_R}.
    \end{align}
We can therefore conclude that in the general case the situation is not significantly different from the simplified model discussed in the main text. In fact, the twistedness can in principle be transferred from the photon to the center of mass of the atom under the realistic experimental conditions.
%Substituting Eq.~\eqref{eq:MS1} into Eq.~\eqref{eq:STW} and then into Eq.~\eqref{eq:wf} of the main text, we get

\section{Transfer of the quantum state of light to the center of mass. General case.}

We first note that equation (21) of the main text is general and does not depend on the particular form of the coefficient $a$. Under the paraxial approximation helicity and projection of the photon spin $\sigma=0,\pm1$ on the $z$ axis are equal $\sigma \approx \Lambda$ and the following approximate decomposition of the polarization vector can be used
\begin{align}
\label{eq:Gpot}
{\bf A}({\bf r}) \approx \boldsymbol{\chi}_{\Lambda} \exp(ik_z z) \int e^{i{\bf{k}_\perp}({\bf r})} a({\bf k}_\perp) \frac{d^2 k_\perp}{(2\pi)^2}.
\end{align}  
Above $\boldsymbol{\chi}_{\Lambda}$ is given by \eqref{eq:chi} and as before $\Lambda=\pm1$.  Coefficients $a({\bf k}_\perp)$ sets the transverse structure of the photon wave function that can be any (Bessel, Hermite, Airy etc.). Now we refer  to the Eq.(25) of the main text that with Eq.\eqref{eq:Gpot} reads 
\begin{eqnarray}
\label{eq:WFTW}
&\Phi_b^{\rm TW}&(t,{\bf R}) = \exp[-iE_bt+i(P_{z,a}-k_z)z]\int e^{i{\bf P}_{\perp,b}\mathbf{R}} a_{\kappa}({\bf P}_{\perp,b}) \frac{d^2P_{\perp,b}}{(2\pi)^2} \nonumber\\
&\times&\delta(E_a+\varepsilon_a+\omega-E_b-\varepsilon_b) \widetilde{\cal M}_{m_b m_a}(\theta_k).
\end{eqnarray}
By comparing Eq.\eqref{eq:Gpot} and Eq.\eqref{eq:WFTW} we immediately observe that the transverse structure of the photon wave function is transferred to the atomic center of mass, as the scalar integral that defines the transverse structure of the wave function is essentially the same.

\end{widetext}

\bibliography{refs}

\end{document}